\documentclass[prb,reprint,amsmath,amssymb,aps]{revtex4-1}
\usepackage{graphicx}
\usepackage{dcolumn}
\usepackage{bm}
\usepackage{amsmath}
\usepackage{bm}
\usepackage{hyperref}

\newcommand{\ks} {{\bf k}}
\newcommand{\vs} {{\bf v}}

\newcommand{\fc} {{\bf F}}

\newcommand{\ps} {{\bf p}}
\newcommand{\ds} {{\bf d}}

\newcommand{\rs} {{\bf r}}

\newcommand{\es} {{\bf e}}
\newcommand{\ac} {{\bf A}}

\newcommand{\gc} {{\bf G}}
\newcommand{\eps} {{\varepsilon}}

\begin{document}

\title{High harmonic generation from bulk diamond driven by intense femtosecond laser pulse}

\author{Tzveta Apostolova$^{1,2}$}

\author{B.  Obreshkov$^1$}

\affiliation{$^1$ Institute for Nuclear Research and Nuclear Energy, 1784 Sofia, Bulgaria}
\affiliation{$^2$ Institute for Advanced Physical Studies, New Bulgarian University, 1618 Sofia, Bulgaria}

\begin{abstract}

We present theoretical results on the high-harmonic generation (HHG) in bulk
diamond induced by intense laser pulse of wavelength 800 nm and duration 15 fs.
For laser intensity in the range $1 \le I \le 50 $ TW/cm$^2$ above bandgap harmonics are
generated after the pulse peak. We find that the intensity of individual harmonics increases non-linearly with the peak laser intensity, 
following a non-perturbative trend.  For moderate intensity the HHG spectrum exhibits a primary plateau with noisy odd order harmonic structure and a cutoff. 
For increased laser intensity a secondary plateau emerges with quasi-continuous spectrum of harmonics extending beyond the 50th order.
Consistently with experimental observations, we find that the cutoff energy for HHG scales linearly with the peak field strength and derivation of the cutoff law is provided.
\end{abstract}

\maketitle

\section{Introduction}

Diamond is a material exhibiting unique mechanical, thermal and electronic properties,
such as high electron and hole mobilities and high breakdown field strength \cite{Iseberg}.
Ultrashort intense laser pulse laser-matter interaction can induce and probe electron dynamics at unprecedented short time scales.
In particular, high-harmonic generation (HHG) is one of the most important phenomena in the strong-field laser-matter interaction.
HHG is a secondary process of non-linear laser frequency conversion to its multiples driven by the
strong coupling of the oscillating electric field of an intense ultrashort laser pulse to electrons.
HHG has been investigated experimentally and theoretically over a broad range of laser intensities and wavelengths,
in  gaseous and molecular targets \cite{Pherson_JOSAB1987,Ferray_JPB1988,Kulander_1993,L'Huillier_PRL1993,Macklin_PRL1993,Christov_PRL_97},
plasmas \cite{Teubner_RevModPhys_2009} and in the solid state \cite{Ghimire_Nphys2011}.
The experiments have shown that sources based on high harmonic generation (HHG) have
spatial and temporal coherence, high directionality and
polarization but due to the
low efficiency of the process, the intensities of the generated
harmonics are orders of magnitude lower than that of the

driving laser field. They are less intense than
EUV,  XUV and soft x-ray radiation produced in the huge facilities such as FEL
accelerator \cite{Elmeguid_Tesla} and Synchrotron Radiation (SR)
\cite{Winick_2013}.  Nevertheless they can be potentially
used to obtain coherent radiation of sub-femtosecond to
attosecond duration in a broad wavelength range (XUV to X-ray)
\cite{Farkas_PLA_1992,Corkum_2007,Krausz_Rev_Mod_Phys_2009,Christov_Opt_Comm_98}
capable of inducing physical processes on very short time and space scales leading to applications such as
attosecond electron dynamics \cite{Schultze_SCI2014}, molecular tomography and
protein crystallography \cite{Chapman_ NAT2011}, etc.

Generation of high harmonics in the gas phase is usually described by a semi-classical three step model \cite{Lewenstein_PRA_1994,Corkum_PRL_1993,Schafer_PRL_1993},
including field ionization, followed by the acceleration of the photoelectron and its subsequent recombination with the parent ion resulting in emission of single energetic photon.
HHG spectra typically consist of sub-bandgap harmonics with rapidly decreasing intensity in agreement with perturbation theory,and
a plateau region of above bandgap harmonics with a high energy cutoff where the harmonic intensity
drops off abruptly. The cutoff energy of the gas-phase harmonics depends linearly on the laser intensity according to the relation
$\hbar \omega_{{\rm max}} \approx I_p + 3.17 U_p$, where $I_p$ is the atomic ionization potential and $U_p$ is the pondermotive energy \cite{Krause_PRL_1992,Watson_PRL_1997}.

Experimental studies of HHG in semiconductors and dielectrics have been reported recently \cite{Ghimire_Nphys2011,Ndabashimiye_Nat2015,Luu_Nat_2015}.
In solid crystals the interplay between generation of carriers via multiphoton
excitation over the bandgap at high intensities leading to
catastrophic breakdown of the material and HHG is carefully
explored to find regimes in which only HHG takes place avoiding
material collapse. In the pioneering study of Ghimire et al. \cite{Ghimire_Nphys2011} long
wavelength laser radiation is used to suppress multiphoton ionization and to obtain HHG.
Experimental and theoretical study of HHG spectra from rare-gas solids \cite{Ndabashimiye_Nat2015}
exposed characteristic features of HHG in solids: emergence of multiple plateaus  as the driving laser intensity
increases over a narrow intensity range that consists of a few harmonics; complicated dependence of the cutoff on the driving laser field,
i.e. following neither linear nor square-root scaling. Furthermore, the experimentally meausred HHG spectra display clean harmonic peaks near multiples of the driving laser frequency.  
This finding has been interpreted as a result of ultrafast polarization dephasing process with dephasing time less than an optical half-cycle \cite{Vampa_PRL2014}. 
More recently, mesoscopic-scale effects inlcuding pulse propagation and inhomogeneous energy depostion in the diamond bulk were shown to lead to clean harmonic spectra 
in good qualitative agreement with the experimental data \cite{Floss_arxiv_2017}.

Various theoretical models have been proposed
to rationalize HHG in crystalline solids including HHG in diamond varying from time-dependent-density-functional-theory,
\cite{Otobe_JCTN2009,Otobe_JAP2012,Otobe_PRB2016,Tancogne-Dejean_PRL2017}
time-dependent-Schrodinger-equation (TDSE) \cite{Higuchi_PRL2014,Wu_PRA2015}, semiconductor Bloch equations \cite{Schubert_NatPhot2014,Luu_PRB2016}
and semiclassical methods \cite{Vampa_cutoff} the latter being more intuitive than the former.
The underlying microscopic processes are still explored to elucidate further the dependence of HHG in solids on the electronic band structure and on the
interplay between inter- and intraband transitions as well as on the collective behavior of the high density electron-hole plasma coupled to the lattice.
In this respect methods based on the TDSE in single-particle approximation incorporating realistic band structure may provide helpful insight on the mechanisms of HHG in solids.

In view of the fact that HHG in wide band-gap semiconductors may lead to brighter sources due to the high laser-induced carrier density
and ability to withstand the driving laser fields necessary for inducing strong nonlinear and non perturbative effects leading to HHG, in the present work
we explore the feasibility for sub-femtosecond pulse generation in diamond bulk subjected to intense 15 fs laser pulse
with 800 nm near-infrared wavelength. We choose near-infrared laser frequencies to stimulate underresonant photoexcitation of electrons 
in diamond and investigate their non-linear response in the tunneling regime (small Keldysh parameter). 
For the fixed pulse duration, the considered laser intensity interval $1 \le I \le 50$ TW/cm$^2$  corresponds to fluences well 
below the experimentally measured threshold for surface ablation of diamond $F_{{\rm th}} \approx $ 6 J/cm$^2$ (cf. Ref. \cite{apostolova_apsus_2018}).

The paper is organized as follows: In Section II the theoretical framework of photoexcitation and generation of high harmonics is presented.
Section III presents numerical results and discussion of HHG spectra. Unless otherwise stated, atomics units ($e=\hbar=m_e=1$) are used throughout this paper. 

\section{Theoretical description}

We describe the electron dynamics in a unit cell of crystalline diamond subjected to spatially uniform laser electric field  $\fc(t)$
with the time-dependent Schr\"{o}dinger equation in single-active electron approximation
\begin{equation}
i \partial_t |\psi_{v\ks} (t) \rangle= H(t)
|\psi_{v\ks}(t) \rangle, \label{tdse}
\end{equation}
where $v \ks$ labels the initially occupied valence band states of
definite crystal momentum $\ks$ and
\begin{equation}
H(t) = \frac{1}{2}[\ps+\ac(t)]^2+V(\rs), \label{ham}
\end{equation}
is the time-dependent Hamiltonian in velocity gauge with $\fc(t)=-\partial_t \ac(t)$, $\ps=-i \nabla_{\rs}$ is the momentum operator,
$V(\rs)$ is an empirical ion-lattice pseudopotential
\begin{equation}
 V(\rs)=\sum_{\gc} V(G) \cos(\gc \cdot \boldsymbol \tau) e^{i \gc \cdot \rs}, \label{vps}
\end{equation}
here 2$\boldsymbol{\tau}=a_0(1/4,1/4,1/4)$ is a relative vector
connecting two carbon atoms in a crystal unit cell, $a_0=3.57$ {\AA} is the bulk lattice constant and $\gc$ labels the
reciprocal lattice wave-vectors.  In Eq.(\ref{vps}), the pseudopotential formfactors $V(G)$ (in Rydberg)
are $V (G^2 = 3) = −0.625, V (G^2 = 8) = 0.051$, and $V (G^2 =
11) = 0.206$; here the wave number $G$ is given in units of $2 \pi/a_0$.
The pseudopotential parameters reproduce the principal energy gaps of the static band
structure of diamond, the indirect bandgap of 5.42 eV and an direct gap of $\Delta=$7.08 eV at the BZ center \cite{epm_diamond1970}. 
To solve Eq.(\ref{tdse}), we expand the time-dependent state-vector over static Bloch orbitals according to 
\begin{equation}
 |\psi_{v \ks}(t) \rangle = \sum_{v'=1}^{N_v} a_{vv'\ks} (t) |v' \ks \rangle + \sum_{c=1}^{N_c} a_{vc\ks} (t) |c \ks \rangle, \label{wp}
\end{equation}
and numerically solve the resulting coupled equations for the coefficients in the Fourier expansion, cf. Ref. \cite{Lagomarsino_PRB2016}.
The bulk band structure is represented by the $N_v=4$ valence bands and $N_c=16$ unoccupied
conduction bands. The Brillouin zone (BZ) was sampled by a Monte Carlo
method using 3000 randomly generated $\ks$-points. The laser vector potential was assumed to be
a temporally Gaussian pulse $\ac(t)=\es \exp(-ln(4) t^2/\tau^2) (F/\omega_L)  \sin \omega_L t$, where $\hbar \omega_L$=1.55 eV is the photon energy,
$\tau=$ 15 fs is the time duration and $\es$ is a unit vector in the direction of laser polarization; the crystal was irradiated by a single pulse
linearly polarized along the $\Lambda$ line ($\Gamma$-L direction).  The time-step for the numerical integration of the equation of motion
was $\delta t=0.03$ a.u.

The optically-induced electron current in the direction of the laser
electric field $\es$ is given by the BZ integral
\begin{equation}
J(t)=\int_{{\rm BZ}} \frac{d^3 \ks}{4 \pi^3} J(\ks,t)
\end{equation}
where
\begin{equation}
\label{j}
J(\ks,t)=\sum_v \langle \psi_{v\ks}(t) | \es \cdot \vs(t) |
  \psi_{v\ks}(t) \rangle ,
\end{equation}
where $\vs(t)=\ps+\ac(t)$ is the velocity operator.  The HHG spectrum is presented as a coherent sum
\begin{equation}
I(\omega)=\left|\int_{{\rm BZ}} \frac{d^3 \ks}{4 \pi^3}
J_{\ks,\omega} \right| \label{hhg_spec}
\end{equation}
over contributions from Bloch states with wave-vector $\ks$
\begin{equation}
J_{\ks,\omega}=\int dt e^{i \omega t} J(\ks,t).
\end{equation}

\section{Numerical Results and discussion of high harmonic spectra}

In Fig.\ref{fig:current_density}(a-c) we plot the time evolution of the velocity distribution of electrons
for  peak laser intensities $I=$ 2,10 and 30 TW/cm$^2$, respectively.
In the low-intensity regime shown in Fig.\ref{fig:current_density}a, the velocity distribution consits of two oppositely directed streams of electrons 
with speeds modulated by the laser vector potential, i.e. 
the distribution follows the quasi-adiabatic time evolution with
\begin{equation}
J(\ks,t) \approx J^{{\rm ad}}(\ks,t) = \left. \sum_v \es \cdot \nabla_{\ks} \eps_v(\ks) \right|_{\ks=\ks(t)}
\end{equation},
where $\ks(t)=\ks+\ac(t)$ is the shifted crystal momentum and $\eps_v(\ks)$ are the dispersion curves for valence electrons.  
For higher intensity $I=$ 10 TW/cm$^2$, Fig.\ref{fig:current_density}b,
the velocities near the BZ edges still follow the
adiabatic time evolution, however states in a narrow strip about the BZ center do not have enough time to fully equilibrate
against the laser vector potential, such that the adiabatic time evolution is disrupted after the peak of the pulse.
As a consequence the non-linear polarization current near the $\Gamma$ point undergoes very
rapid and irregular sub-cycle changes resulting in generation of high harmonics after the pulse peak. In the high-intensity
regime with $I=$ 30 TW/cm$^2$, as shown in Fig.\ref{fig:current_density}c,  the velocity distribution
is strongly distorted at the BZ center and this non-adiabatic distortion extends towards the BZ edges.

The optically induced currents are shown in Fig. \ref{fig:current_tot}(a-c) for the three different laser intensities discussed above.
Stable and reversible (free from breakdown) ultrafast currents following the laser vector potential are induced, cf. Fig. \ref{fig:current_tot}(a-c).
For the increased laser intensity, rapid sub-cycle oscillations of the current become noticeable in the wake of the pulse, cf. Fig. \ref{fig:current_tot}c. 
To dipslay the relatively weak non-linear response of the diamond crystal, we split the total current into linear and non-linear contributions 
\begin{equation}
J(t) = J_{{\rm lin}}(t) + J_{{\rm nlin}}(t),   
\end{equation}
where the linear response  
\begin{equation}
 J_{{\rm lin}}(t) = \frac{1}{\pi} \Re \int_0^{2 \omega_L} d \omega e^{-i \omega t} J(\omega) ,
\end{equation}
is obtained by inverse Fourier transformation of the total current; the non-linear currents are shown in Fig. \ref{fig:current_tot}(d-f). 
In the low intensity regime in Fig. \ref{fig:current_tot}d, the transient current displays the generation of an intense third harmonic superimposed 
on the 5th bandgap harmonic. The undamped harmonic oscillations of the current after the end of the pulse with period corresponding to
the bandgap energy $\Delta$ is a direct consequence of the build-up of coherent superposition of populations between the valence and the conduction band.
For the increased laser intensity as shown in Fig.\ref{fig:current_tot}e, the rapid and irregular sub-cycle changes of the current become prominent. 
After the pulse is over, the optically-induced current exhibits a complicated beating pattern due to $\Gamma_{25'} \rightarrow \Gamma_{2'}$ (7 eV)
and $\Gamma_{25'} \rightarrow \Gamma_{15}$ (10 eV) interband transitions, the dominant oscillation frequency (0.4 fs) corresponds to the 
$\Gamma_{25'} \rightarrow \Gamma_{15}$ transition. Similar result is found for the highest laser intensity shown in Fig. \ref{fig:current_tot}f,
which makes evident the relevance of strong interband couplings among  multiple conduction band states. The convergence of the current with 
respect to the number of $\ks$-points used in the numerical integration  is shown in Fig. \ref{fig:hhg_kpts}(a-b);  
the two curves present comparison of calculations emplying 3000 and 5000 quasimomentum points, respectively. As this comparsion demonstrates, denser sampling of BZ reduces  
the amplitude of the AC current Fig.\ref{fig:hhg_kpts}, which in turn makes more pronounced the non-linear effects, cf. Fig.\ref{fig:hhg_kpts}b. 
Aside from the overall amplitude decrease, we also find weak attosecond time shift between the two current waveforms. Though the amplitude of the current 
is reduced with the increase of the number of $\ks$-points, we find that the non-linear response and HHG can be adequately represented 
by using a smaller set of 3000 points.  

The spectral line strength $I(\ks,\omega)=\left| J_{\ks\omega} +J_{-\ks,\omega}\right|$ for the $\Lambda$ line is also shown in Fig.\ref{fig:hhg_kw}(a-c) .
For relatively weak and moderate peak laser intensity, Fig.\ref{fig:hhg_kw}(a-b), the odd harmonic structure of the velocity distribution is clearly exhibited.
For the lowest intensity shown in Fig.\ref{fig:hhg_kw}(a), sub-bandgap harmonics
extend from the BZ center towards the edges. The rapid sub-cycle distortion of the velocities
near the BZ center gives rise to intense 5th harmonic peak, the 7th order is
highly suppressed. For the increased laser intensity with $I=$ 10 TW/cm$^2$ shown in Fig.\ref{fig:hhg_kw}b, the direction-dependent HHG
spectrum along the $\Lambda$-line is characterized by a plateau region which terminates near and above the position of the 13th harmonic.
The lack of even order harmonics is because for monochromatic laser irradiation with $A(t)=(F/\omega_L) \sin \omega_L t$,
the time-dependent Hamiltonian in Eq.(\ref{ham}) is invariant under the combined transformation of spatial inversion $\rs \rightarrow -\rs$
supplemented by discerete time translation $t \rightarrow t+ m \pi/\omega_L$ for $m$ odd-integer, so that 
\begin{equation}
J(\ks,t)=-J \left(-\ks,t+ m \frac{\pi}{\omega_L} \right), \quad m=1,3,\ldots    \label{dyns}
\end{equation}
Because of this dynamical symmetry, the coefficients in the Fourier series in $t$: $J(\ks,t)=\sum_n J_n(\ks)\exp(i n \omega_L t)$,
satisfy the relation $J_n(\ks)=-(-1)^n J_n(-\ks)$, so that all even order harmonics interfere destructively and do not contribute to the total HHG current.
Though the symmetry relation implied by Eq.(\ref{dyns}) is explicitly broken by the finite pulse duration $\tau$, 
the lack of even order harmonics remains prominent in Fig.\ref{fig:hhg_kw}(a-b). For the highest intensity shown in Fig.\ref{fig:hhg_kw}c, 
the spectrum extends to high orders and the odd-harmonic structure is smeared out.
The primary frequency comb extends to 15th order, decreases gradually in spectral intensity with the increase of the harmonic
order and crosses over into a secondary plateau with harmonics extending beyond the 50th order.

In Fig. \ref{fig:hhg_spec}(a-c) we show the coherently summed HHG spectra as a function of the harmonic order for the three peak laser intensities $I=2,10$ and 30 TW/cm$^2$.
In the low intensity regime, Fig. \ref{fig:hhg_spec}a, the spectrum is characterized by clear and pronunced peaks at odd harmonic orders. 
The intensity of the 5th harmonic is enhanced relative to 3rd one, due to dynamical Stark shift, cf. Ref.\cite{Lagomarsino_PRB2016}; 
the spectrum exhibits a cutoff regime with fast fall-off for photon energies $> 5 \hbar \omega_L$, such that for relatively weak laser 
field the high-energy cutoff is  determined primary by the direct bandgap energy at the $\Gamma$ point. For the increased laser intensity shown in Fig.\ref{fig:hhg_spec}b,
the spectrum extends to higher orders with primary plateau region and a cut-off regime above the position of 9th harmonic; though harmonic orders 
up the 21th are distinct, the intensity of the frequency comb apparently exhibits fast decline above the 9th order. Indeed such a fall-off above the 9th order is not apparent in the 
crystal orientation-dependent line strength shown in Fig. \ref{fig:hhg_kw}b, which makes evident that HHG in diamond bulk is anisotropic, cf. also Ref. \cite{You_nphys_2017}. 
Furthermore the discrete character of individual harmonics in the plateau regime is blurred; this lack of clear odd-harmonic structure can be rationalized 
within the semi-classical re-collision model as due interferences of multiple re-collisions between electron-hole pairs with different total energies 
at different times of recombination \cite{Vampa_cutoff}. For the highest peak intensity $I=$ 30 TW/cm$^2$ shown in Fig.\ref{fig:hhg_spec}(c), 
HHG is characterized by quasi-continuous spectrum of harmonics and emergence of weak secondary plateau with cutoff positioned at the  50th harmonic. 
This secondary plateau with harmonic orders extending beyond the atomic limit of the
corresponding gas-phase harmonics indicates strong dependence of HHG on the band structure of diamond;
Fig.\ref{fig:hhg_2} demonstrates that when the total number of conduction states in Eq.(\ref{wp}) is reduced from $N_c=$ 16 to 1,
the secondary plateau is truncated, which makes evident that this plateau originates in the recombination of electrons promoted highly in the conduction band
with valence band holes. Similar results are found in HHG spectra from rare-gas solids \cite{Ndabashimiye_Nat2015}.

In Fig.\ref{fig:si_F} we plot the field dependence of the spectral intensity of individual harmonics $S_n$
in the first plateau region; here $n$ designates the harmonic order. 
The peak laser intensity dependence can be divided into 3 regimes. I.
Perturbative regime:  The 3rd and the 5th harmonic are distinguishable and scaling law  $S_n \sim F^n$ is  exhibited. A crossing point is observed, 
where an inversion in the  efficiencies of the 3rd and 5th harmonics occurs. 
II. Intermediate regime (with $F \sim 0.1$ V/{\AA): The HHG spectrum
extends to higher orders as the field strength increases. The
intensity of different harmonic orders exhibit the power law $
S_n \sim F^6$ regardless on the harmonic order. III. High field
strength with $F \ge 0.5$ V/{\AA}: the intensity of individual harmonics become indistinguishable.

Fig.\ref{fig:cutoff}(a-b) shows the field-dependence of the photon-energy cutoff for the primary and secondary plateau,
respectively. In contrast to the cutoff law observed for gas-phase harmonics,
cutoff harmonics generated from diamond bulk display linear scaling with the field strength $F$.
This linear scaling law is consistent with experimental observations
in ZnO \cite{Ghimire_Nphys2011}. The high-energy cutoff  of the secondary plateau is distinguished
for strong fields with $F \sim$ 0.4 V/{\AA}  corresponding to laser intensity $I > 10$ TW/cm$^2$.
The slope of this linear dependence implies that the cutoff energy increases by 2 harmonic orders when the field strength
is increased by 0.5 V/nm. In contrast, the slope of the photon-energy cutoff of the primary plateau is flatter by a factor
of 3.  The linear scaling law is a consequence of the specific dispersion law between energy and momentum for Bloch electrons 
(cf. also Appendix -Derivation of the cutoff law). For localized (in momentum space) excitation of charge carriers  near the BZ center 
with wave-vector $\ks \approx {\bf 0}$, the total energy of an electron-hole pair moving in the laser field can be written in the form (cf. Ref.\cite{Keldysh})
$\Delta(\ks(t))=\Delta \sqrt{1+ \ks^2(t)/ m \Delta}$,  where $m$ is the reduced mass of the pair.
Within the classical re-collision model \cite{Vampa_cutoff}, an electron-hole pair is born at rest near the peak of the laser field due to tunnel ionization,
the field accelerates away the photo-excited carriers, which then recombine at a later time as the field reverses direction.
The maximal energy gained at the time of recombination is $E_{{\rm max}} \approx \Delta \sqrt{1+ 1/ \gamma_K^2}$,
where $\gamma_K=\omega_L \sqrt{m \Delta}/F$ is the Keldysh parameter, so that for strong laser fields with $\gamma_K \ll 1$ (as in the present work), 
we have $E_{{\rm max}} \approx \Delta/\gamma_K$ which heuristically explains the linear scaling law of the cutoff energy for HHG.

\section{Conclusion}

In summary, we presented calculation of HHG spectrum from bulk diamond induced
by an intense femtosecond laser pulse. Our result describes the
intense HHG extending from the perturbative to highly
non-perturbative incident laser intensity regimes. The
non-perutrbative regime of HHG occurs for laser intensity $I \ge $
1 TW/cm$^2$. We find that depending on the laser intensity,
the HHG spectra from bulk diamond exhibits two
plateau regions with above bandgap harmonics, the secondary plateau extends beyond the limit of the
corresponding gas-phase harmonics emitted for the same laser
intensity. The appearance of multiple plateaus turns out to be specific for HHG in
solids characterized by strong interband couplings between multiple conduction band states.
The linear scaling law of cutoff energy for HHG is sensitive to the concrete type of
energy-momentum dispersion law for Bloch electrons. The numerical results make  evident the feasibility of diamond for attosecond pulse generation.

\section*{Acknowledgements}
This work is supported by the Bulgarian National Science Fund under Contract No.  DFNI-E02/6 and Contract No. DNTS/FRANCE-01/9 (T.A.) and by
the Bulgarian National Science Fund under Contract No. 08-17 (B.O.).

\appendix*
\subsection*{Appendix: Derivation of the cut-off law} 

Because the disruption of the velocity distribution of photoexcited carriers in diamond is localzied in momentum space about the $\Gamma$ point, 
and since the asymptotic velocity distribution includes primary contributions of electron-hole pairs moving in $\Gamma$ bands, 
a derivation of the cutoff energy for HHG can be based on a simplified two-band model with the Hamiltonian 
\begin{equation}
H_{{\rm eff}}(\ks,t)=\frac{\Delta}{2}(1+\tau_3)+V_{\ks}(t) \tau_1, \label{heff} 
\end{equation}
where $(\tau_1,\tau_2,\tau_3)$ are the three Pauli matrices, and the coupling among the valence and conduction band is defined by  
\begin{equation}
V^2_{\ks}(t)=\sum_{i=1}^3 [\ks(t) \cdot \ps_{cv_i}({\bf 0})]^2 ,
\end{equation} 
which takes into account the three-fold degeneracy of the valence-band at the $\Gamma$ point $\ks=(0,0,0)$, here $\ps_{cv}({\bf 0})$ 
is the matrix element of the momentum operator at the BZ center. Similarly to the standard $\ks \cdot \ps$-method \cite{Cardona}, 
the Hamiltonian in Eq.(\ref{heff}) adequately represents  the static band structure of diamond near the $\Gamma$ point; the interaction with the 
laser field is introduced by the substitution $\ks \rightarrow \ks(t)=\ks+\ac(t)$; while $\ks$ is small, the shifted momentum $\ks(t)$ may not be small.  
The time-dependent Schr\"{o}dinger equation
\begin{equation}
i |\partial_t \psi_{\ks}(t) \rangle=H_{{\rm eff}}(\ks,t) |\psi_{\ks}(t)\rangle 
\end{equation}
is subject to the initial condition
\begin{equation}
|\psi_{\ks}(0)\rangle = |-(\ks(0))\rangle ,
\end{equation}
where we defined the instantaneous eigenvectors of $H_{{\rm eff}}(t)$ 
\begin{equation}
|+(\ks(t))\rangle=\left( 
\begin{array}{c} 
\cos(\theta_{\ks}/2) \\
\sin(\theta_{\ks}/2)
\end{array}
\right), 
\quad  
|-(\ks(t))\rangle=\left( 
\begin{array}{c} 
-\sin (\theta_{\ks}/2) \\
\cos (\theta_{\ks}/2)
\end{array}
\right)
\end{equation}
with eigen-energies 
\begin{equation}
 \eps_{\pm}(\ks(t))=\frac{\Delta}{2} \pm \sqrt{\Delta^2/4 + V^2_{\ks(t)}} 
\end{equation}
and a mixing angle by the definition $\cos \theta_{\ks}(t)=\Delta/\sqrt{\Delta^2 + 4 V_{\ks}^2(t)}$. Expanding the state-vector in this adiabatic basis
$|\psi_{\ks}(t) \rangle = c_{\ks}(t) |+(\ks(t))\rangle + d_{\ks}(t) |-(\ks(t))\rangle$,
and treating the non-adiabatic coupling $ \tau_2 \dot \theta_{\ks} $ as  weak, to first order in perturbation theory 
the total current  can be written as
\begin{eqnarray}
\label{jtot}
& & J(t)= \sum_{\ks} \ps_{vc}(\ks(t))  \int_{-\infty}^t dt'  \fc(t') \cdot \ds_{cv}(\ks(t')) \\
& & e^{-i S(\ks,t,t')} + c.c.
\end{eqnarray}
where $\ds_{cv}(\ks)=i \ps_{cv}(\ks)/\Delta_{\ks}$ is the transition dipole moment, 
$\Delta_{\ks}=\eps_+(\ks)-\eps_-(\ks)=\sqrt{\Delta^2+4 V_{\ks}^2}$ is the momentum-dependent transition frequency and
\begin{equation}
S(\ks,t,t')=\int_{t'}^t dt'' \Delta_{\ks}(t'')
\end{equation}
is the classical action for the motion of an electron-hole pair in the laser field. Eq.(\ref{jtot}) has a simple and intuitive interpretation 
which corresponds to the semi-classical three-step model of HHG in the gas phase \cite{Lewenstein_PRA_1994,Corkum_PRL_1993}: at time $t'$  the factor $\fc(t') \cdot \ds_{cv}(\ks(t'))$ 
describes the photo-ionization process, while the factor $\ps_{vc}(\ks(t))$  determines the amplitude for recombination of an electron-hole pair at time $t$; 
the phase factor $\exp(-i S(\ks,t,t'))$  describes the propagation of the charge-carriers from $t'$ to $t$. At a given recombination time $t$, the photocurrent is obtained 
by integration over the contributions from all ionization times $t'<t$ and all crystal momenta $\ks$. Exploiting the cubic symmetry of the diamond lattice, we approximate 
the matrix element $\langle c {\bf 0} | p_a p_b | c {\bf 0} \rangle \approx \frac{\sigma^2}{3} \delta_{ab}$, where $\sigma^2 = \langle c {\bf 0} |\ps^2 | c {\bf 0} \rangle = 2.76$ 
is the mean-squared fluctuation of the carrier momentum at the BZ center, such that the energy-momentum dispersion law exhibits a relativistic form  
$\Delta_{\ks}  \approx \sqrt{\Delta^2+4 \sigma^2 k^2/3}$. Formal equivalence with the classical relativistic dynamics of charged particle in electromagnetic field
can be established by introducing the notation $4\sigma^2/3=c^2$ and the reduced mass of an electron-hole pair $m=\Delta/c^2$, so that
\begin{equation}
 \Delta_{\ks}(t)=c \sqrt{m^2 c^2 + \ks^2(t)}, 
\end{equation}
and the Keldysh parameter is $\gamma_K=m c \omega_L/F$.

Similarly to the gas phase, the cutoff law of solid state harmonics can be derived from classical saddle point analysis, 
the saddle point equations are
\begin{eqnarray*}
& & \nabla_{\ks} S(\ks,t,t')= \int_{t'}^t d t'' \vs(t'')  = {\bf 0} \\
& & \partial_{t'} S(\ks,t,t')=\Delta_{\ks}(t')=0 \\
& & \partial_t S(\ks,t,t')=\Delta_{\ks}(t)=\hbar \omega \\,
\end{eqnarray*}
here $\vs(t)=\ks(t)/\Delta_{\ks}(t)$ is the group velocity of the wave-packet. The first equation states that HHG occurs, when electron and hole pair re
-collide after being accelerated and split apart by the laser field after the time of ionization. The second equation specifies the complex time of ionization, 
and the third equation expresses the energy conservation law, i.e. the electron-hole pair recombines radiatively at time $t$  by emitting a photon of energy $\hbar \omega$.

To derive the cut-off law for HHG, the acceleration of the electron-hole pair subjected to monochrmatic laser field $F(t)=F \cos \omega_L t$ is
\begin{equation}
 \frac{d \vs}{dt}=\frac{F \cos \omega_L t}{m \gamma(v)}(\es - \frac{1}{c^2}\vs (\vs \cdot \es))
\end{equation}
where $\gamma(v)=1/\sqrt{1-\beta^2}$ is the Lorentz factor with $\beta=v/c$. This equation is subject to the initial condition $\vs(t_i)={\bf 0}$, 
i.e. we assume that an electron-hole pair is born at rest at the time of ionization $t_i=t'$, such that it is sufficient to study the dynamics in 
the direction of the laser polarization, setting $\es=(0,0,1)$, we have $v_x=v_y=0$ and $v_z=v$, i.e.
\begin{equation}
\frac{d v}{dt}=\frac{F \cos \omega_L t}{m \gamma^3(v)},
\end{equation}
the relevant quantity is the kinetic energy gain 
\begin{equation}
E_{{\rm kin}}= (\gamma(v)-1) \Delta
\end{equation}
For a given time of ionization $t_i$, the time of recombination $t_r=t$ is obtained from the solution of the equation 
$z(t_r,t_i)=0$, here $z(t,t_i)=\int_{t_i}^t dt' v(t')$ is the separation between the electron and the hole at time $t$. 
The energy of a photon emitted upon recombination is $\gamma(v(t_r)) \Delta$, the maximal energy at the time of recombination 
specifies the high-energy cutoff 
\begin{equation}
\hbar \omega_{{\rm max}} = \Delta \max_{t_i} \gamma(v(t_r(t_i)))
\end{equation}
Using the definition
\begin{equation}
\beta(t) \gamma(t) = \frac{1}{\gamma_K}(\sin \omega t -\sin \omega t_i ),
\end{equation}
we observe that the speed passes through zero $\beta(t_{\ast})=0$  at the time
$t_{\ast}=\pi/\omega_L-t_i$. This reflection occurs for all trajectories with times of ionization less than one half-period of the laser field $t_i < \pi /\omega_L$. 
In the high speed limit with $|\beta| \approx 1$ (or equivalently $\gamma_K \ll 1$) we use the approximation
$\beta(t) \approx \tanh \omega_L(t-t_{\ast})/\gamma_K$ to model the rapid change of speed at $t_{\ast}$. 
The time of recombination is determined from $\int_{t_i}^{t_r} dt \beta(t) = 0$,  which results in
\begin{equation}
2 t_{\ast} -(t_i+t_r) \approx 0,
\end{equation}
i.e. $t_r=2 \pi/\omega_L - 3 t_i$, so that the Lorentz factor at the time of recombination becomes
\begin{equation}
\gamma_r=\frac{1}{\gamma_K} |\sin \omega_L t_i + \sin 3 \omega_L t_i|, 
\end{equation}
and at the same time the total energy of the charge carriers is  
\begin{equation}
E(t_r)=\gamma_r \Delta= \frac{\Delta}{\gamma_K} |\sin \omega_L t_i + \sin 3 \omega_L t_i|. 
\end{equation}
The trigonometric factor $|\sin x + \sin 3 x|$ is maximized for $x \approx  \pi/5$, and therefore the maximal photon energy upon recombination is
\begin{equation}
\label{eq:maxE}
\hbar \omega_{{\rm max}} \approx 1.54 \frac{\Delta}{\gamma_K}=1.54 \frac{F}{\omega_L} \left(\frac{\Delta}{m}\right)^{1/2}
\end{equation}
which suggests that the cutoff energy scales linearly with the laser wavelength and is in qualitative agreement with the experimentally observed 
linear scaling law of the cutoff energy for HHG in widebandgap solids.

\begin{figure}
\begin{center}
\includegraphics[width=.5\textwidth]
{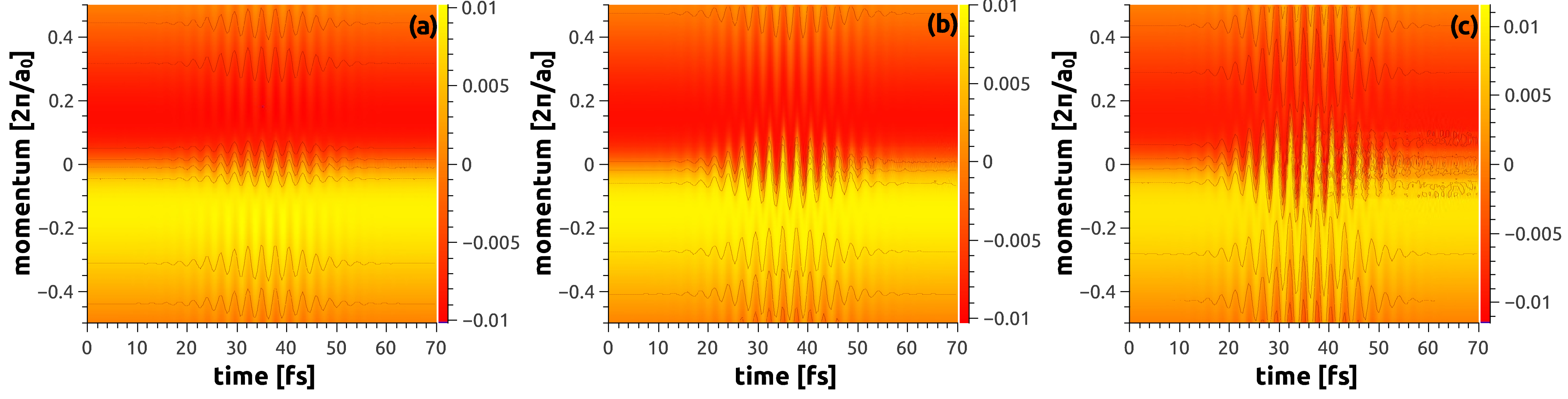} \caption{Velocity distribution of electrons along the $\Lambda$-line in diamond bulk driven by intense laser pulse with wavelength 800 nm and time duration 15 fs.
The peak laser intensity is: (a) I=2 TW/cm$^2$, (b) I=10 TW/cm$^2$ and (c) I=30 TW/cm$^2$ }
\label{fig:current_density}
\end{center}
\end{figure}

\begin{figure}
\begin{center}
\includegraphics[width=.5\textwidth]
{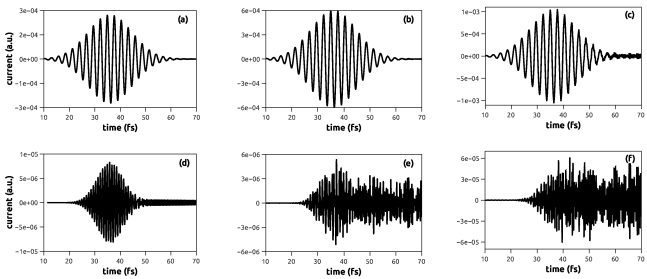} \caption{Currents in diamond bulk driven by an intense laser pulse with wavelength 800 nm and time duration 15fs.
The peak laser intensity is: (a) I=2 TW/cm$^2$, (b) I=10 TW/cm$^2$ and (c) I=30 TW/cm$^2$. 
For the same laser intensities in (a-c), the non-linear part of the current is shown in Figs. (d-f), respectively.}
\label{fig:current_tot}
\end{center}
\end{figure}

\begin{figure}
\begin{center}
\includegraphics[width=.5\textwidth]
{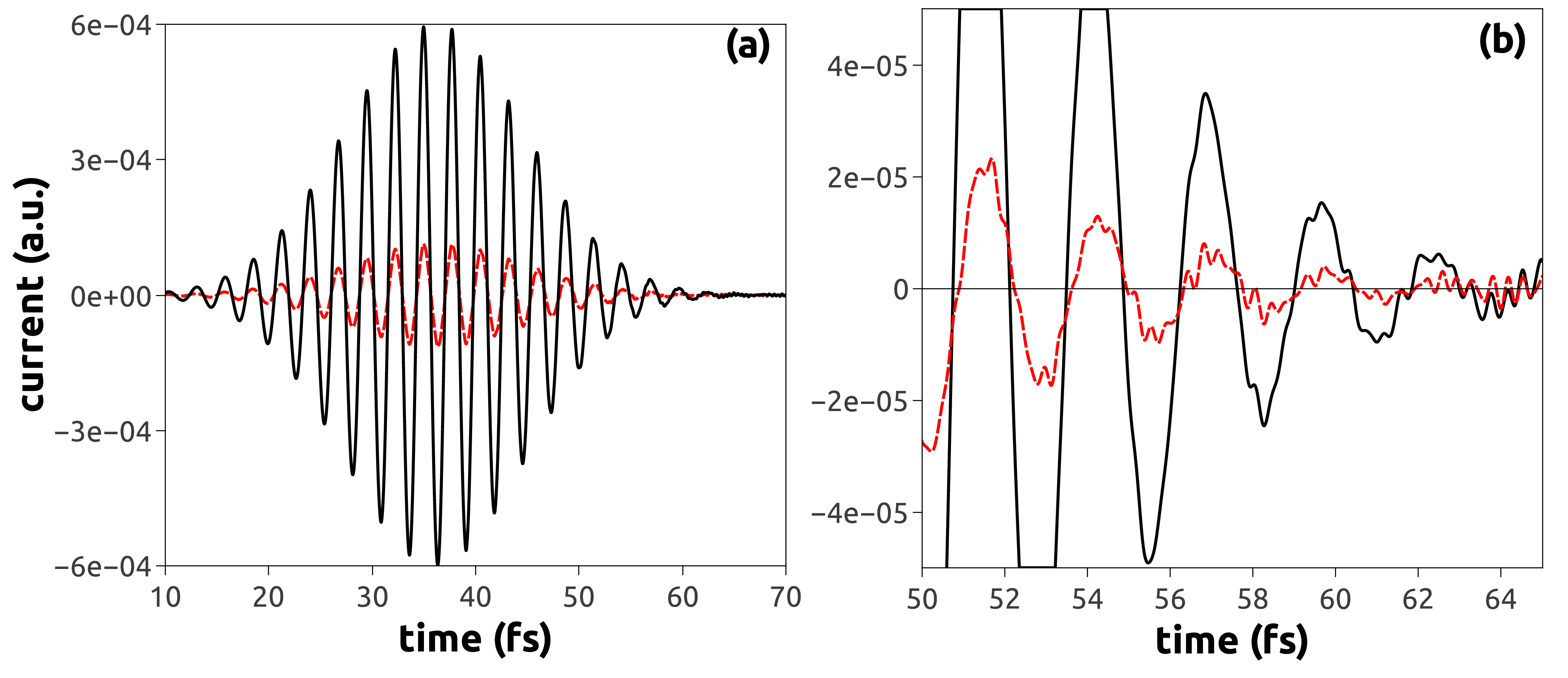} \caption{ Convergence of the optical currents in diamond bulk with respect to the number of $\ks$-points in the Brillouin zone.
The laser wavelength is 800 nm, the pulse duration is 15 fs and the laser intensity is 10 TW/cm$^2$. 
The red dotted-line in Fig(a-b) is a result of calculation employing 5000 $\ks$-points, the black solid line in employs 3000 points.}
\label{fig:hhg_kpts}
\end{center}
\end{figure}

\begin{figure}
\begin{center}
\includegraphics[width=.5\textwidth]
{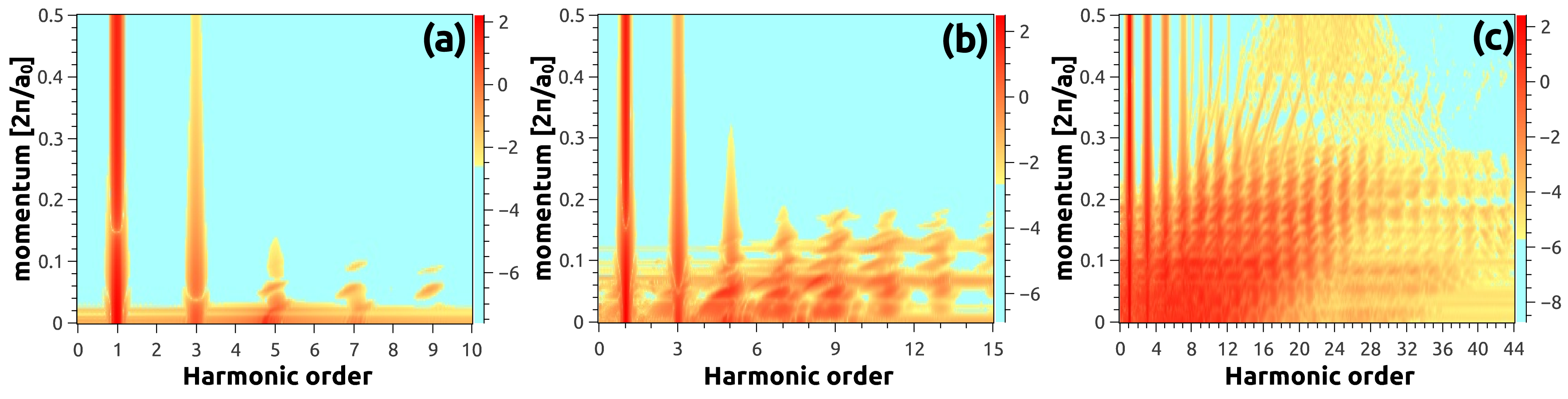} \caption{Fourier transform of the velocity distribution of electrons along the $\Lambda$ line in the Brillouin
zone. The peak laser intensity is: (a) I=2 TW/cm$^2$, (b) I=10 TW/cm$^2$ and (c) I=30 TW/cm$^2$}
\label{fig:hhg_kw}
\end{center}
\end{figure}

\begin{figure}
\begin{center}
\includegraphics[width=.5\textwidth]
{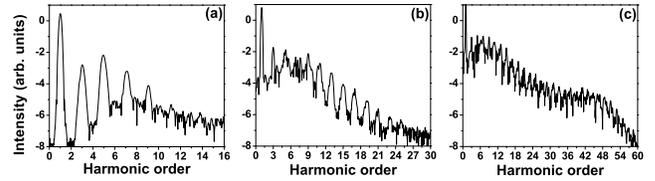} \caption{HHG spectra obtained for photon energy 1.55 eV and pulse duration 15 fs. The
peak laser intensity is 2,10 and 30 TW/cm$^2$}
\label{fig:hhg_spec}
\end{center}
\end{figure}

\begin{figure}
\begin{center}
\includegraphics[width=.3\textwidth]
{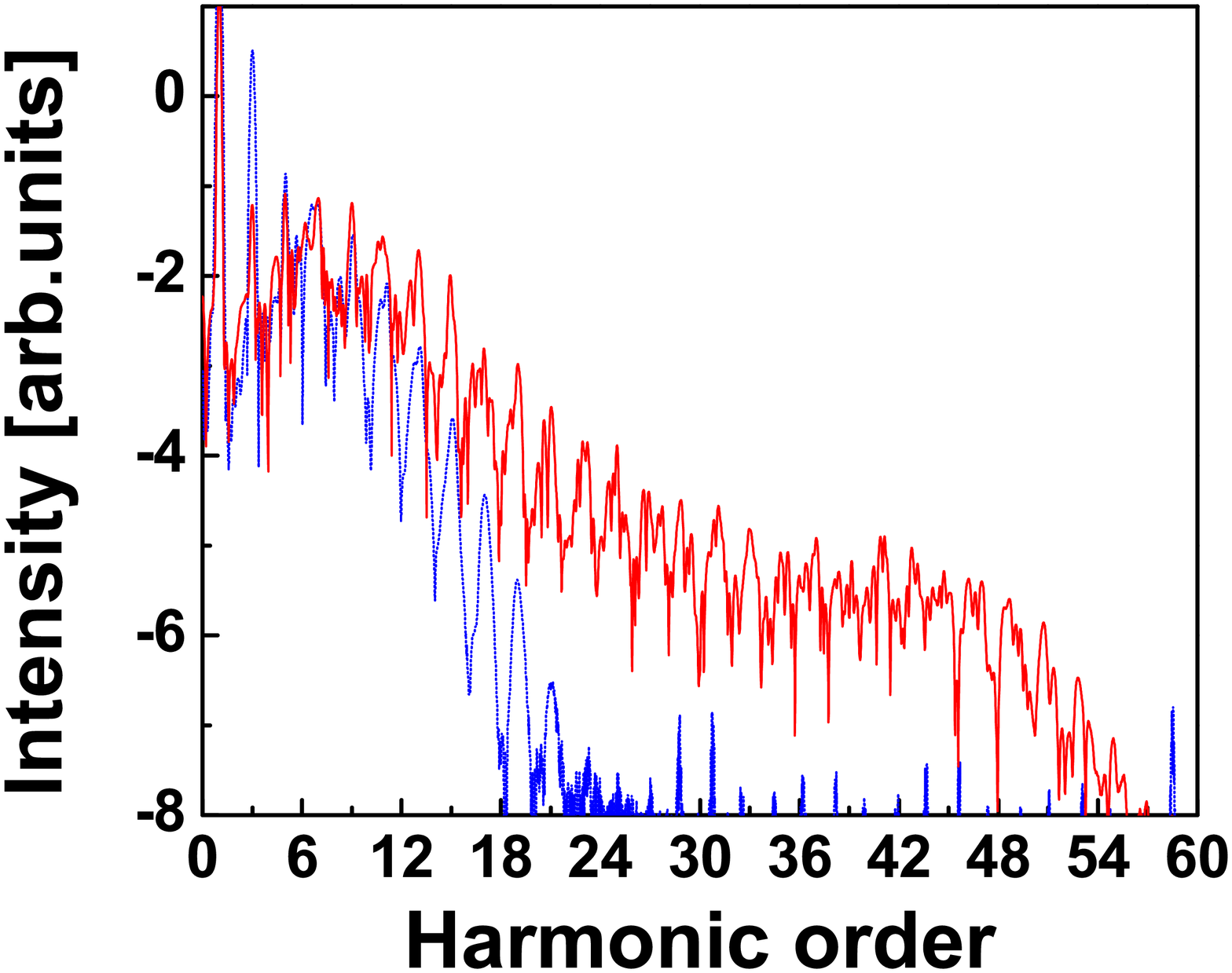} \caption{HHG spectra obtained for photon energy 1.55 eV, pulse duration 15 fs and
intensity 30 TW/cm$^2$. The number of conduction bands is $N_c=1$ (dashed curve) and $N_c=16$ (solid curve)}
\label{fig:hhg_2}
\end{center}
\end{figure}

\begin{figure}
\begin{center}
\includegraphics[width=.5\textwidth]
{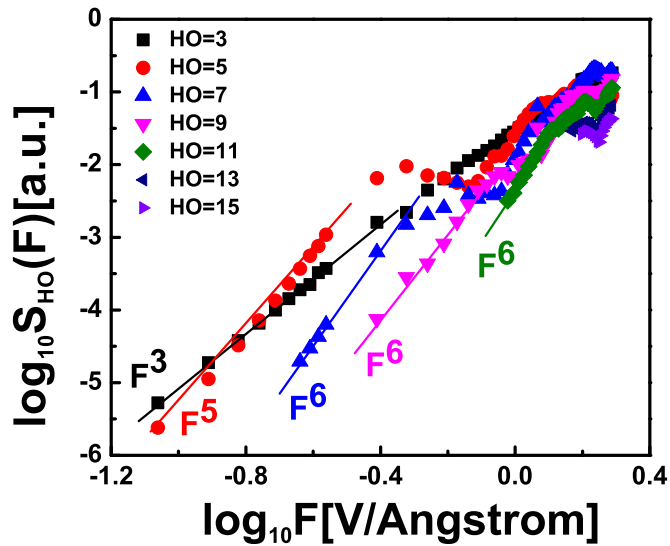} \caption{Dependence of the spectral intensity of individual harmonics on the electric field strength
for 15fs laser pulse with photon energy 1.55 eV interacting with bulk diamond} \label{fig:si_F}
\end{center}
\end{figure}

\begin{figure}
\begin{center}
\includegraphics[width=.5\textwidth]
{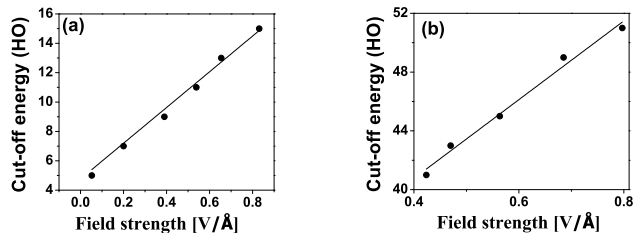} \caption{ High-energy cutoff as a function of the driving laser field for
the primary plateau in Fig.(a) and the secondary plateau in Fig.(b)} \label{fig:cutoff}
\end{center}
\end{figure}






\end{document}